\documentclass[sigconf,screen]{acmart}

\usepackage{multirow}
\usepackage{adjustbox}
\usepackage{hyperref}
\usepackage[capitalize]{cleveref}
\usepackage{enumitem}
\usepackage{microtype}
\DisableLigatures[-]{family=tt*}
\usepackage{dblfloatfix}    

\usepackage{listings}
\AtBeginDocument{%
  \providecommand\BibTeX{{%
    \normalfont B\kern-0.5em{\scshape i\kern-0.25em b}\kern-0.8em\TeX}}}
\acmYear{2022}\copyrightyear{2022}
\setcopyright{acmlicensed}
\acmConference[MMSys '22]{13th ACM Multimedia Systems Conference}{June 14--17, 2022}{Athlone, Ireland}
\acmBooktitle{13th ACM Multimedia Systems Conference (MMSys '22), June 14--17, 2022, Athlone, Ireland}
\acmPrice{15.00}
\acmDOI{10.1145/3524273.3532890}
\acmISBN{978-1-4503-9283-9/22/06}

\newcommand{\ie}[0]{\textit{i.e.},}
\newcommand{\eg}[0]{\textit{e.g.},}

\usepackage{array,graphicx}
\usepackage{booktabs}
\usepackage{textcomp}

\newcommand{\mO}[0]{\checkmark{}} 
\newcommand{\mRes}[2]{{\scriptsize\tt #1 \(\times\) #2}} 
\newcommand{\mNov}[2]{\the\numexpr #1 * #2 \relax\,\,{\scriptsize(\(=\!#1\!\times\!#2\))}} 
\newcommand{\mVarious}[0]{{\it \small various}} 
\newcommand{\mVariousMP}[1]{\mVarious{}\;{\scriptsize(\(\pm #1\)MP)}} 

\newcommand{\mZ}[1]{\O\;(#1)}
\newcommand{\mT}[0]{tiny}
\newcommand{\mS}[0]{small}
\newcommand{\mM}[0]{medium}
\newcommand{\mL}[0]{large}

\newcommand*\rot{\rotatebox{90}}
\newcolumntype{R}[2]{%
    >{\adjustbox{angle=#1,lap=\width-(#2)}\bgroup}%
    c%
    <{\egroup}%
}
\newcommand\rotd[1]{\multicolumn{1}{R{45}{1em}}{\small #1}}

\newcommand{\bettershortstack}[2][c]{%
  \begingroup\renewcommand{\arraystretch}{0.7}\begin{tabular}[b]{@{}#1@{}}#2\end{tabular}\endgroup%
}

\begin{document}

\newcommand{\dsname}{SILVR} 
\newcommand{\dsnameexpl}{\textit{\textbf{S}ynthetic \textbf{I}mmersive \textbf{L}arge-\textbf{V}olume \textbf{R}ay} dataset} 
\newcommand{\gardenname}{Zen Garden}
\title{\dsname{}: A Synthetic Immersive Large-Volume Plenoptic Dataset}

\author{Martijn Courteaux}
\email{Martijn.Courteaux@UGent.be}
\orcid{0000-0002-9971-3128}
\affiliation{%
  \institution{Ghent University -- imec}
  \streetaddress{Technologiepark-Zwijnaarde 122}
  \city{Zwijnaarde}
  \state{Oost-Vlaanderen}
  \country{Belgium}
  \postcode{9052}
}

\author{Julie Artois}
\email{Julie.Artois@UGent.be}
\orcid{0000-0003-2822-5183}
\affiliation{%
  \institution{Ghent University -- imec}
  \streetaddress{Technologiepark-Zwijnaarde 122}
  \city{Zwijnaarde}
  \state{Oost-Vlaanderen}
  \country{Belgium}
  \postcode{9052}
}

\author{Stijn De Pauw}
\email{Stijn.DePauw1@gmail.com}
\orcid{0000-0002-0158-602}
\affiliation{%
  \institution{Ghent University -- imec}
  \streetaddress{Technologiepark-Zwijnaarde 122}
  \city{Zwijnaarde}
  \state{Oost-Vlaanderen}
  \country{Belgium}
  \postcode{9052}
}

\author{Peter Lambert}
\email{Peter.Lambert@UGent.be}
\orcid{0000-0001-5313-4158}
\affiliation{%
  \institution{Ghent University -- imec}
  \streetaddress{Technologiepark-Zwijnaarde 122}
  \city{Zwijnaarde}
  \state{Oost-Vlaanderen}
  \country{Belgium}
  \postcode{9052}
}

\author{Glenn Van Wallendael}
\email{Glenn.VanWallendael@UGent.be}
\orcid{0000-0001-9530-3466}
\affiliation{%
  \institution{Ghent University -- imec}
  \streetaddress{Technologiepark-Zwijnaarde 122}
  \city{Zwijnaarde}
  \state{Oost-Vlaanderen}
  \country{Belgium}
  \postcode{9052}
}

\renewcommand{\shortauthors}{Courteaux, et al.}

\begin{abstract}
In six-degrees-of-freedom light-field (LF) experiences, the viewer's freedom is limited by the extent to which the plenoptic function was sampled.
Existing LF datasets represent only small portions of the plenoptic function, such that they either cover a small volume, or they have limited field of view.
Therefore, we propose a new LF image dataset ``\textit{\dsname{}}'' that allows for six-degrees-of-freedom navigation in much larger volumes while maintaining full panoramic field of view.
We rendered three different virtual scenes in various configurations, where the number of views ranges from 642 to 2226.
One of these scenes (called \textit{\gardenname{}}) is a novel scene, and is made publicly available.
We chose to position the virtual cameras closely together in large cuboid and spherical organisations (\(2.2\text{m}^3\) to \(48\text{m}^3\)), equipped with 180\textdegree{} fish-eye lenses.
Every view is rendered to a color image and depth map of \(2048\text{px}\times2048\text{px}\).
Additionally, we present the software used to automate the multi-view rendering process, as well as a lens-reprojection tool that converts between images with panoramic or fish-eye projection to a standard rectilinear (\ie{} perspective) projection.
Finally, we demonstrate how the proposed dataset and software can be used to evaluate LF coding/rendering techniques (in this case for training NeRFs with instant-ngp).
As such, we provide the first publicly-available LF dataset for large volumes of light with full panoramic field of view.

\end{abstract}




\maketitle

\section{Introduction and Motivation}




In today's multimedia landscape, photos and videos are omnipresent and a lot of research and development activities are targeting more immersive ways to consume content.
Using technologies such as Virtual Reality (VR), one can navigate with six-degrees-of-freedom (6DoF) through a virtual environment.
This has led to a surge in research to methodologies for capturing and representing real-world environments, such that these can be be streamed and/or displayed in real time with 6DoF.
Many such methodologies and systems start from images or videos of this environment as their source of data.
The combination of many individual photographic captures of an environment, with accurate knowledge of the used lens and camera position for every image, is called a light field (LF) capture.

Examples that demonstrate the potential of using LF captures over traditional single-camera captures include Google's ``Welcome To Light Fields'' \cite{welcome_to_lightfields} and DeepView \cite {deepview} and Meta's ``Manifold in Action'' \cite{pozo2019manifold}.
These allow for high-quality, real-time rendering in VR, but the freedom of motion of the viewer is confined by a small sphere determined by the used capture setup.
Enabling research for immersive 6DoF experiences with less confined freedom of movement relies on the availability of datasets that capture the environment in an immersive way over larger volumes.
Unfortunately, to the best of our knowledge, no currently available dataset has the combination of properties required for large-volume 6DoF immersive navigation.
For this reason, we propose a novel dataset ``\textit{\dsname{}}'' (\dsnameexpl{}) with the following features:
\begin{enumerate}[label=(\roman*)]
    \item
        Three photo-realistic synthetic scenes, rendered with ground-truth depth maps.
        The scenes contain several elements that are known to pose challenges for many LF technologies, such as specularities, volumetrics, reflections, and translucency.
    \item
        High-resolution (\(2048\text{px}\times2048\text{px}\)) fish-eye (180\textdegree{} field of view) cameras are evenly distributed ($\pm0.1$m apart) across the surface of a cuboid or sphere.
        This leads to a dense sampling of the plenoptic function, and thus of the light rays entering the cuboid or sphere.
    \item
        The cuboids and spheres are large (\(2.2\text{m}^3\) to \(48\text{m}^3\)), with cameras looking outward.
        This allows for 6DoF in a large volume.
\end{enumerate}

The dataset consists of static images, ground-truth depth maps, and configuration files containing details about the intrinsics and extrinsics of each camera.
In addition to this, we make two tools openly available: the addon that was developed to render the images with Blender, and a reprojection tool that converts between panoramic or ultra-wide and rectilinear images.
The dataset and tools can be found on \url{https://idlabmedia.github.io/large-lightfields-dataset}, together with more in-depth information, examples and instructions on how to reproduce all results.

\section{Related Datasets}\label{sec:sota}


\begin{table*}[ht]
\small
\centering
\begin{tabular}{cl|cccc|cccccc}
    &
    & \multicolumn{4}{c}{Environment} & \multicolumn{6}{|c}{Data properties} \\
    & Dataset &
    \rotd{Natural / Synthetic} &
    \rotd{Specularities} &
    \rotd{\bettershortstack[l]{Volumetrics /\\ Translucencies}} &
    \rotd{Dynamic} &
    \rotd{Resolution} &
    \rotd{Interpolation Volume} &
    \rotd{\bettershortstack[l]{Immersive\\(Omni-directional/360\textdegree{})}} &
    \rotd{Has depth} &
    \rotd{\bettershortstack[l]{Nr. of views per scene}} &
    \rotd{Note}
    \\
    \midrule
\multirow{8}*{\rot{\scriptsize Inside-out LFs}}
& Stanford LF Archive \cite{NewStanfordArchive}       & N   & \mO  &      &     & \mVariousMP{1}    & \mT      &      &         & \mNov{17}{17}  & (1)  \\
& OrangeKitchen \cite{mpeg_datasets}                  & S   & \mO  &  \mO & \mO & \mRes{1920}{1080} & \mT      &      &  \mO    & 25             &       \\
& NokiaChess \cite{mpeg_datasets}                     & S   & \mO  &      & \mO & \mRes{2048}{2048} & \mS      & \mO  &  \mO    & 10             &       \\
& Technicolor LF Dataset \cite{Sabater2017}           & N   &      &      & \mO & \mRes{2048}{1088} & \mS      &      &         & \mNov{4}{4}    &       \\
& Google Spaces \cite{googlespaces}                   & N   & \mO  & \mO  &     & \mRes{2048}{1229} & \mS      &      &         & 16             &       \\
& \citeauthor{pozo2019manifold}
  6DoF Video Camera \cite{pozo2019manifold}           & N   &      &      & \mO & \mRes{3160}{2160} & \mM      & \mO  &         & 16             & (2)   \\
& \citeauthor{4DLFVD} 4DLFVD \cite{4DLFVD}            & N   & \mO  & \mO  & \mO & \mRes{1920}{1056} & \mS      &      &         & 100            &       \\
& \citeauthor{broxton2019lowcostcamera}
\cite{broxton2019lowcostcamera}                       & N   & \mO  & \mO  & \mO & \mRes{2560}{1920} & \mM      &      &         & 46             & (3)   \\
\midrule
\multirow{4}*{\rot{\scriptsize Outside-in LFs}}
& Meta CO3D \cite{reizenstein21co3d}                  & N   & \mO  & \mO  &     & \mVarious         & -        &      &         & \mVarious      &       \\
& Google RealEstate10K \cite{realestate10k}           & N   & \mO  & \mO  &     & \mVarious         & -        &      &         & \mVarious      & (4)   \\
& DTU Multi View Stereo \cite{aanaes2016large}        & N   & \mO  &      &     & \mRes{1600}{1200} & -        &      &         & 49 or 64       &       \\
& NeRF (llff) data \cite{mildenhall2020nerf}          & N   & \mO  & \mO  &     & \mRes{4032}{3024} & -        &      &         & 20 to 62       &       \\
\midrule
\multirow{4}*{\rot{\scriptsize Multi 360\textdegree}}
& ClassroomVideo \cite{mpeg_datasets}                 & S   &      &      & \mO & \mRes{4096}{2048} & \mZ{\mM} & \mO  & \mO     & 15             &     \\
& Matterport3D \cite{Matterport3D}                    & N   & \mO  & \mO  &     & \mRes{1280}{1024} & \mZ{\mM} & \mO  & \mO     & \(18 n\)       & (5) \\
& \citeauthor{armeni2017joint}
2D-3D-semantic \cite{armeni2017joint}
                                                      & N   & \mO  & \mO  &     & \mRes{1280}{1024} & \mZ{\mM} & \mO  & \mO     & \(18 n\)       & (5) \\
& \citeauthor{FTV360} FTV360 \cite{FTV360}            & N   &      &      & \mO & \mRes{3840}{1920} & \mZ{\mL} & \mO  &         & 40             &     \\
\midrule
\multirow{3}*{\rot{\scriptsize \textbf{\textit{Ours}}}}
& \textbf{\textit{Barbershop}}                        & S   & \mO  & \mO  &     & \mRes{2048}{2048} & \mL      & \mO  & \mO     & 642 or 1400    &    \\
& \textbf{\textit{Lone Monk}}                         & S   & \mO  &      &     & \mRes{2048}{2048} & \mL      & \mO  & \mO     & 642 or 2226    &    \\
& \textbf{\textit{\gardenname{}}}                     & S   & \mO  & \mO  &     & \mRes{2048}{2048} & \mL      & \mO  & \mO     & 642 or 1806    &    \\
        \bottomrule
    \end{tabular}
    \vspace{0.1em}
    \caption{Overview of existing scenes and datasets related to LF or immersive media. Notes:
    (1)~Tarot Cards has interesting light refraction through a glass ball.
    (2)~16 Fisheye lenses on a sphere with diameter 1m, yielding an IIV of approximately \(0.52\text{m}^3\).
    (3)~46 Fisheye lenses on a dome.
    (4)~YouTube videos solved for camera intrinsics and extrinsics, meaning mostly 1-dimensional camera traces.
    (5)~For one panoramic image, 18 pictures are taken by a rotating camera on a tripod. The dataset consists of 10800 panoramas for Matterport3D and 1413 for 2D-3D semantic.}
    \label{tab:sota}
\end{table*}

This section discusses some existing relevant LF datasets and assesses their applicability to large-volume immersive 6DoF experiences.
Aspects such as the presence of depth maps, as well as the presence of optical elements such as specularities, volumetrics (\eg{} fog, water, fire, smoke), and translucency are considered.
\Cref{tab:sota} summarizes the datasets described below.
Most datasets fall in one of three categories: \textit{inside-out light fields}, \textit{outside-in light fields}, \textit{multiple panoramic 360\textdegree{} captures}.
\citeauthor{broxton2020immersive} introduced the term \textit{interpolation volume} (IV)~\cite{broxton2020immersive} (or \textit{ray intersection volume} in \cite{broxton2019lowcostcamera}) as the volume behind the cameras that consists of the points for which incoming light rays can be obtained by interpolating data captured from the cameras~\cite{broxton2020immersive,broxton2019lowcostcamera}.
We extend this concept and introduce \textit{immersive interpolation volume} (IIV), which is the interpolation volume limited to the points for which light rays in \textit{any} direction can be obtained by interpolation.
We are interested in datasets with a large IIV, as these enable immersive 6DoF experiences.

The first category of LF datasets uses cameras in an \textit{inside-out} fashion to capture the environment.
Many datasets in this category use camera setups organized on a plane \cite{NewStanfordArchive,Sabater2017,googlespaces,4DLFVD,mpeg_datasets} or curved surface.
A notable example is the camera rig by \citeauthor{broxton2019lowcostcamera}, where 46 video cameras are mounted on an acrylic dome, which provides an IV which is around 80cm in width~\cite{broxton2019lowcostcamera}.
While some datasets have a considerable IV, they all have an empty IIV, as the cameras were all pointing towards the same region of the environment.
\citeauthor{pozo2019manifold} built a 16-camera rig with cameras positioned on a sphere pointing outward, yielding a spherical IIV with a diameter of approximately 1m, which they use to record immersive video~\cite{pozo2019manifold}.
\citeauthor{deepview} released their \textit{Welcome to Light Fields} software demo on SteamVR along with several LF captures with a spherical IIV with a diameter of approximately 60cm~\cite{deepview,welcome_to_lightfields}.
However, they use a customized closed-source VP9 video codec to compress the LF data~\cite{welcome_to_lightfields}, which makes it difficult to use it for research.
Additionally, LFs captured with plenoptic cameras (\eg{} Lytro Illum) like the \textit{Kalantari} dataset~\cite{LearningViewSynthesis} and \textit{The Stanford Lytro Light Field Archive}~\cite{StanfordLytroLFArchive} also belong in this category, but because they are not immersive (\ie{} no 360\textdegree{} light information is available) and have a tiny IV (only a few centimeters wide), they are not included in \cref{tab:sota}.

The next category of LF datasets employs  \textit{outside-in} capturing for object in an environment, without explicitly capturing the environment itself.
The cameras are positioned around the object and point inward to capture this object.
As such, these LF captures are not considered immersive in the context of this paper.
While the IV is typically large as it reaches entirely around the object, the IIV is empty.
Examples of such datasets include \cite{reizenstein21co3d,realestate10k,aanaes2016large,mildenhall2020nerf} which we will not further discuss but are included in \cref{tab:sota} instead.

The last category of LF datasets uses multiple 360\textdegree{}-panoramic captures from an environment.
These 360\textdegree{} photographs are typically taken on the same elevation from the ground and relatively far apart (\(\pm\)2m), such as the work by \citeauthor{Matterport3D}~\cite{Matterport3D} and \citeauthor{armeni2017joint}~\cite{armeni2017joint} in which the \textit{Matterport} camera is used.
The Matterport camera is put on a rotating tripod and produces 18 images to cover the full horizontal 360\textdegree{} field of view.
However, the poles (above and below the camera) are missing from the field of view, and the resulting images are of rather low resolution (\(1280\text{px} \times 1024\text{px}\)), which likely results in noticeable aliasing in modern VR headsets.
The FTV360 dataset by \citeauthor{FTV360} captures videos from 40 tripod-mounted 360\textdegree{} cameras simultaneously~\cite{FTV360}.
The distance between neighboring cameras is between 1m and 3m and camera positions are estimated for all recorded sequences.
The fact that the different viewpoints are far apart makes such datasets unsuitable for LF interpolation techniques, as these typically require denser sampled data.
We reflect this shortcoming in \cref{tab:sota} by assigning an empty IV (\O) for such datasets, although the size of the area they span with the camera positions is included between parenthesis for comparison.
Techniques that do use such sparse data typically rely on accurate depth estimations, and use intermediate representations like meshes or point clouds.
While these techniques deliver immersive 6DoF experiences, they rarely use LF concepts at their core.
Additionally, when capturing an environment sparsely like this, challenging visual elements like specularities, volumetrics, and refraction effects are not adequately captured in the dataset.

We conclude that only the dataset from \citeauthor{pozo2019manifold}~\cite{pozo2019manifold} is truly immersive.
However, while the IIV is 1m in diameter (which is among the largest of real-world camera setups), it consists of only 16 input views.
This low number of views makes it difficult to adequately capture specularities and other difficult elements, as explained for the multiple-360\textdegree{} category.
Overall, there are no public datasets capturing an environment in an inside-out way with a dense camera setup.

\section{Dataset Generation and Description}
\label{sec:dataset}

To address the absence of dense immersive light field datasets, this paper proposes a new dataset with a large interpolation volume covered by many cameras in an immersive configuration.
To this end, three virtual scenes were rendered with Blender's ray-tracing engine \textit{Cycles}.
Section~\ref{sec:dataset-scenes} discusses the scenes content-wise, and Section~\ref{sec:camsetup} describes the selected camera configurations, and the rendering process.

\subsection{Virtual Scenes}
\label{sec:dataset-scenes}
Virtual scenes created in Blender were chosen over real-world environments.
This decision was made for several reasons.
First, rendering software enables a flexible configuration of camera setups and artistic choices, while still being able to produce photo-realistic results.
Second, software lenses are perfect, do not require calibration, and do not require the camera positions and orientations to be known.
Third, rendering can be done on existing general-purpose computer hardware.
Finally, rendering is relatively inexpensive\footnote{With a total rendering time of around 12 days on a Linux system with two NVIDIA Quadro P5000 GPUs, consuming an average 500W (based on the on-device display), total energy usage is estimated to be around 144kWh. Associated electricity costs are estimated to be around 46 euro.}.

Two of the scenes we selected are openly-available demo files from the Blender website\footnote{\url{https://www.blender.org/download/demo-files/}}: \textit{``Agent 327: Barbershop''} (CC-BY, Blender Foundation~\cite{BlenderStudioBarbershop}) and \textit{``Lone Monk''} (CC-BY, by Carlo Bergonzini from Monorender).
Both scenes have enough geometry and decorations to make for an interesting immersive experience.
This is in contrast to artists that only focus on the area that is visible by their chosen camera viewpoint.
The Barbershop scene features the interior of a barbershop with many small props, fine detail, and a mirror.
Not a single patch in this room looks not decorated.
Lone Monk features a courtyard of a cloister with a water well in the center.
Small adjustments were made to make the scene slightly more immersive, as one side of the courtyard had some issues with the roof (these issues were not visible from the original camera).
Additionally, a \textit{solidify} modifier was added to the roof tiles to give them some thickness.

The third scene \textit{``\gardenname{}''} was built from scratch by the authors, using CC-0 models and textures.
We integrated elements which are known to be challenging for LF technologies.
More specifically, the statue and floor stones have reflective materials causing complex specularities.
Additionally, there is a small mirror next to the statue.
There is also a fire, several candles, and lanterns with a translucent case, all of which produce ambiguous depths.
There is fog visible in front of the distant mountains and, finally, there are lots of small details such as tree branches, grass, and bamboo.

\begin{figure*}[ht]
  \centering
  \begin{minipage}[b]{0.32\textwidth}
    \includegraphics[width=\textwidth]{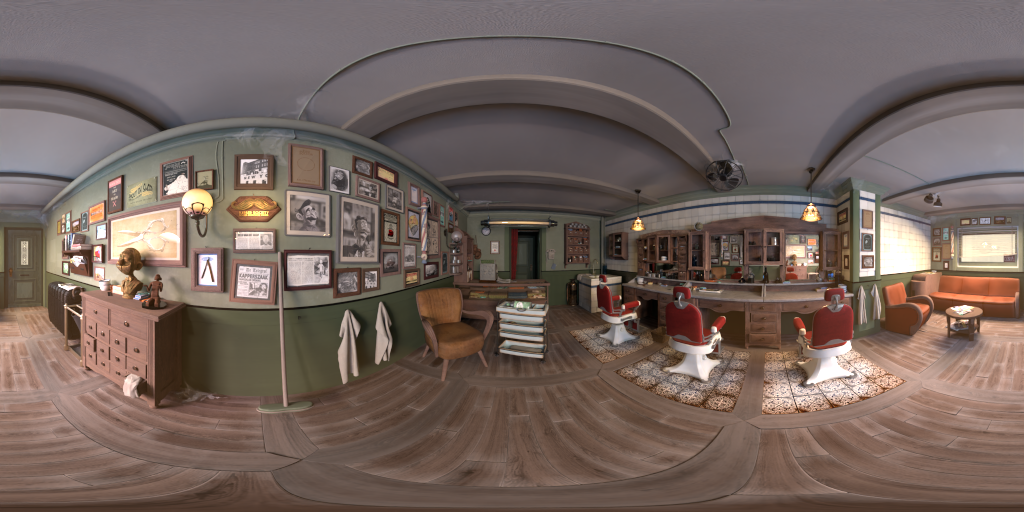}
  \end{minipage}
  \begin{minipage}[b]{0.32\textwidth}
    \includegraphics[width=\textwidth]{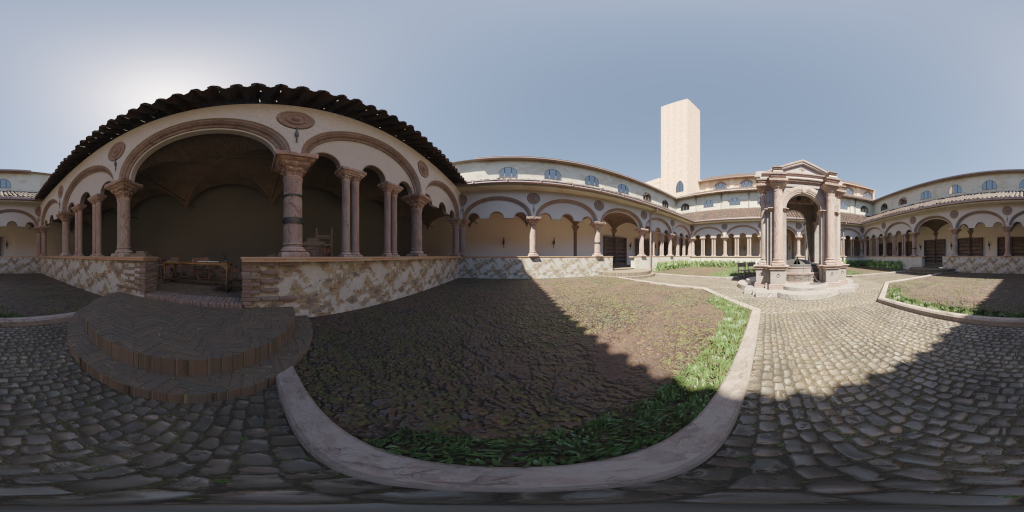}
  \end{minipage}
  \begin{minipage}[b]{0.32\textwidth}
    \includegraphics[width=\textwidth]{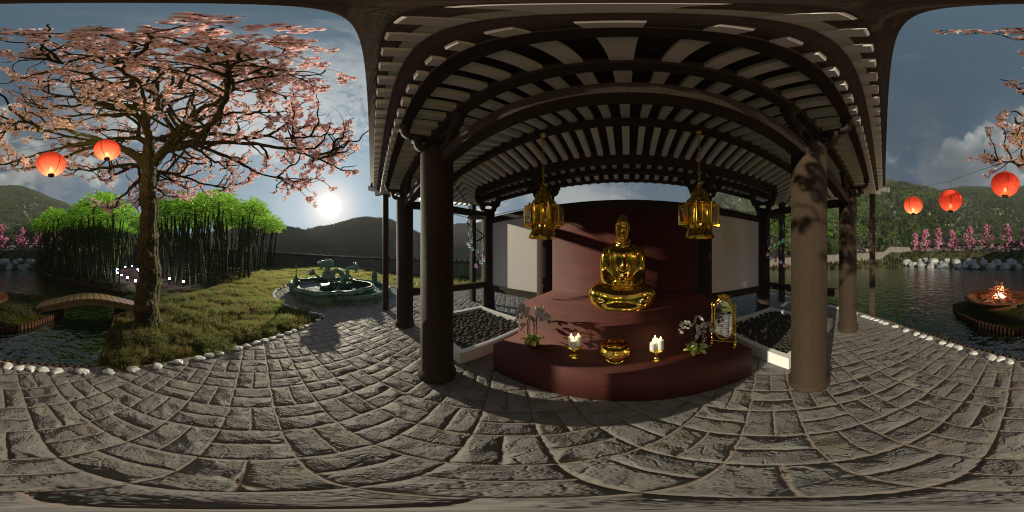}
  \end{minipage}
  \caption{Overview of the Barbershop, Lone Monk, and \gardenname{} scenes.}
  \Description{These three pictures show the scenes Barbershop, Lone Monk and Zen Garden by means of one panoramic 360-degree image each.}
  \label{fig:barbershop_garden_pano}
\end{figure*}

The three rendered scenes are shown in \cref{fig:barbershop_garden_pano}.

\begin{figure}[b]
  \centering
  \begin{minipage}[b]{0.23\textwidth}
    \includegraphics[width=\textwidth]{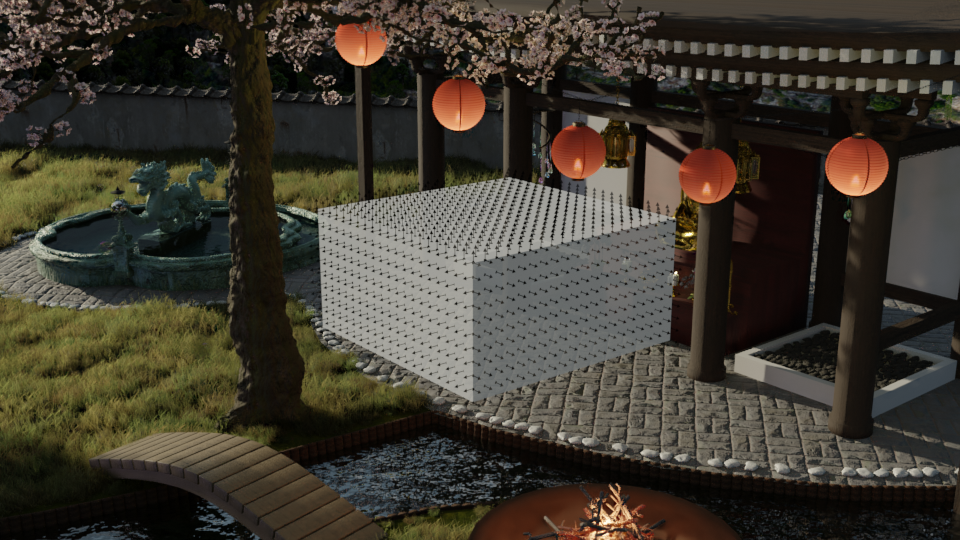}
  \end{minipage}
  \begin{minipage}[b]{0.23\textwidth}
    \includegraphics[width=\textwidth]{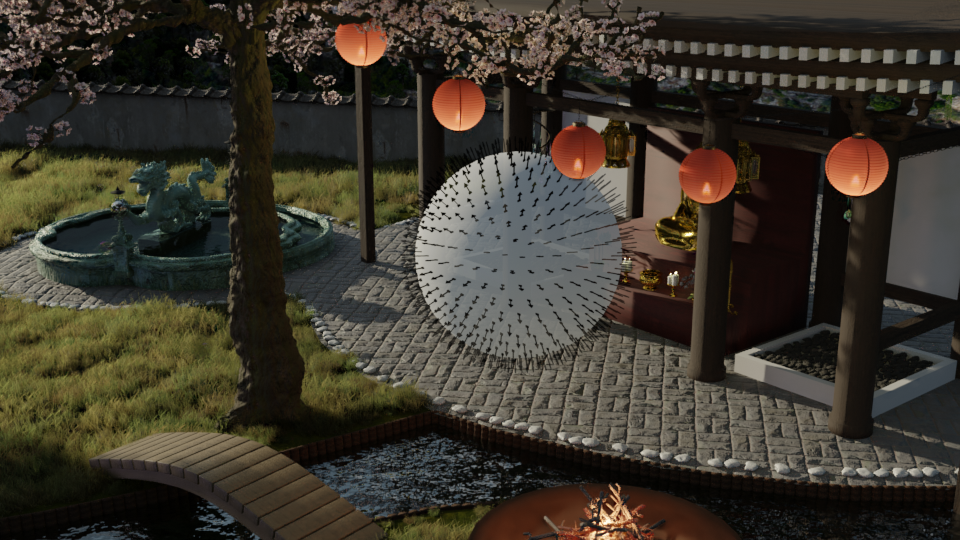}
  \end{minipage}
  \caption{The placement of the cuboid and sphere camera setups in \gardenname{}. The small black arrows indicate the direction each camera is facing. For the cuboid, all cameras belonging to the same face have the same rotation.}
  \Description{These two pictures visualize the camera positions of the two setups (being cuboid and spherical) for the Zen Garden scene that were rendered and provided in this dataset}.
  \label{fig:garden_cuboid_sphere}
\end{figure}

\subsection{Camera Setup and Image Generation}\label{sec:camsetup}
In order to maximize the IIV for a given camera setup, (equidistant) fish-eye lenses with a 180\textdegree{} field of view are used.
We chose to provide two camera-positioning setups per scene: one cuboid and one spherical, as demonstrated in \cref{fig:garden_cuboid_sphere}.
The spherical setup is an elegant way to capture lots light rays from many directions entering the volume.
The cuboid option is added to allow for cameras in six planar setups, which can yield even larger IIVs, as cuboids usually fit easier in closed areas, compared to a sphere.
The specifics of these setups can be found in \cref{tab:camsetups}.
In the table, it can be seen that the IIV is indeed significantly larger than any of the discussed datasets in \cref{sec:sota}, and that the number of views is high to maintain dense spatial sampling.
Additionally, we rendered two extra evaluation configurations: a spherical setup with fewer cameras using perspective lenses, and another setup with 8 equirectangular panoramic renders as seen from the corners of a cube.
These additional renders are provided to serve as ground truth data to evaluate view synthesis algorithms, as they do not coincide with any of the views in the other configurations.
However, the fact that the viewpoints in the base configurations are so numerous and densely positioned enables researchers to subdivide them into input (\eg{} training) and evaluation sets if necessary.
If a uniform division is desirable, we suggest to use the views with odd sequence numbers for evaluation.

\begin{table*}[ht]
\centering
\small
\begin{tabular}{llcccc}
 &  & \multicolumn{2}{c}{Number of views} & \multirow{2}{*}{Dimensions} & \multirow{2}{*}{\bettershortstack{Mean distance\\between cameras}} \\
 &  & in total & along (X, Y, Z) &  &  \\ \midrule
\multirow{2}*{Barbershop}
    & cuboid & 1400 & (10, 30, 10) & 1m \(\times\) 3m \(\times\) 1m & 11cm \\
    & sphere &  642 & -             & 1.45m diameter                 & 9cm \\ \midrule
\multirow{2}*{Lone Monk}
    & cuboid & 2226 & (21, 21, 16) & 4m \(\times\) 4m \(\times\) 3m & 20cm  \\
    & sphere &  642 & -            & 4.0m diameter                  & 31cm \\ \midrule
\multirow{2}*{\gardenname{}}
    & cuboid & 1806 & (21, 21, 11) & 2m \(\times\) 2m \(\times\) 1m & 10cm \\
    & sphere &  642 & -            & 1.7m diameter                  & 13cm \\
 \bottomrule
\end{tabular}
\caption{A summary of the camera setups used in the proposed dataset. For the dimensions of the cuboids, the third value indicates the vertical axis. The mean distance is calculated to each camera's nearest (non-overlapping) neighbour.}
\label{tab:camsetups}
\end{table*}

In order to configure these camera setups easily and render all of the respective views, a novel Blender addon was developed.
This addon is available online at \url{https://github.com/IDLabMedia/blender-lightfield-addon}.
All resulting image files have a resolution of \(2048\text{px}\times 2048\text{px}\) with a circular projection and are stored losslessly in the OpenEXR file format, using 16-bit floating points for all color channels and the depth.
Depth is measured in meters from the camera projection plane (meaning Z-depth, and not Euclidean distance).
For each setup, all camera intrinsics and extrinsics are recorded in a JSON file.

\section{Reprojecting to different lens types}\label{sec:postprocess}

As this dataset provides rendered views with fish-eye projections, these views are not directly suited to for applications requiring perspective projection views.
To address this, we have developed an additional software tool that can convert images between different projections, assuming perfect lenses.
This tool reads the EXR files, reprojects them according to the specified desired lens type and lens parameters, and writes the result back to EXR or PNG format.
It should be noted that only EXR output format will also reproject the depth map, since PNG has no support for accurately storing floating point data required for depth values.
Additionally, when using this tool, one should consider the artifacts of the reprojection process, such as aliasing and upsampling.
Aliasing can be desirable in certain situations, but can be prevented by using multi-sampled reprojection (which can be configured through the command line argument \texttt{\small --samples}).
Perspective projections stretch out the edges of the field of view enormously when using a large field of view.
This can cause upsampling (\ie{} significantly more pixels cover a patch of the field of view in the perspective-projected image than were available in the fish-eye source data).
We therefore suggest to either work with the fish-eye source data directly whenever possible, or consider using smaller field of views (which significantly shrinks the IIV).
Alternatively, to avoid upsampling with large field of views, the result can be downsampled by using the command line argument \texttt{\small --scale} with an argument smaller than 1, which we recommend to combine with multi sampling to avoid aliasing in the center.

\section{Example Use Case: NeRF}
In this Section, we describe the process of preparing the dataset for use in an example application, in this case NVLabs' Instant Neural Graphics Primitives (instant-ngp) implementation for training Neural Radiance Fields (NerF) \cite{mueller2022instant}.
While not fundamental to NeRFs themselves, instant-ngp expects images to use a perspective projection.
As such, in the following example command, we convert the spherical camera setup of Lone Monk to a rectilinear lens (\ie{} lens with perspective projection) with focal length of 18mm, recorded on an image sensor of size 36mm~\(\times\)~36mm, stored as PNG files:

\noindent\begin{minipage}{\linewidth}
\begin{lstlisting}
./reproject --parallel 4 --rectilinear 18,36 --scale 0.125 \
  --png --exposure -1 --reinhard 5 \
  --input-dir lone_monk/LFSphere_e220cm_d400cm/exr \
  --input-cfg lone_monk/LFSphere_e220cm_d400cm/lightfield.json \
  --output-dir lone_monk_perspective \
  --output-cfg lone_monk_perspective/lightfield.json
\end{lstlisting}
\end{minipage}

In the command, the lens type and specifics are extracted from the \texttt{\small lightfield.json} input configuration file.
Note that the dimensions of the image are reduced to 1/8th (\ie{} \(256\text{px}\times256\text{px}\)) and no anti-aliasing measures are taken (\ie{} there is no \texttt{\small --samples 8} flag).
This exploits the fact that instant-ngp can train the NeRF by only generating rays from the pixel centers and that we have a lot of different views available.
Combining many aliased views still yields good results as the aliasing per view is high but the aliasing in the light field in its entirety is relatively low.
Additionally, the EXR files store HDR content, and the Lone Monk scene is too bright to convert to PNG (which does not have HDR support) without adjusting exposure.
As such, the exposure is reduced by one stop ({\tt\small --exposure -1}) and Reinhard tone mapping with maximum brightness 5 ({\tt\small --reinhard 5}) is applied~\cite{reinhard2002}.

Next, the configuration format for instant-ngp is different and uses a different coordinate system.
In the following example command, the generated configuration file from the previous step (that describes the reprojected dataset) is converted to the format of instant-ngp:

\noindent\begin{minipage}{\linewidth}
\begin{lstlisting}
python3 generate_NERF_transforms.py \
  --scene lone_monk \
  --dataset-config lone_monk_perspective/lightfield.json \
  --output-transforms lone_monk_perspective/transforms.json
\end{lstlisting}
\end{minipage}

The used Python script is also available on our GitHub repository.
Note that the `\texttt{\small --scene lone\_monk}' flag is used to use reasonable default values for the scaling and positioning of the scene within the NeRF bounding volume.
After executing the two provided example commands, the generated `\texttt{\small lone\_monk\_perspective}' folder is ready to be opened by instant-ngp%
\footnote{A few animated results can be found at \url{https://idlabmedia.github.io/large-lightfields-dataset}.}.

\section{Conclusion}
In this paper, we presented a new light field dataset designed for six-degrees-of-freedom navigation with full panoramic vision in a large volume.
Due to the large volume, objects can be visible, then occluded, and then visible again when moving linearly through the volume, as the viewpoint can fully traverse past an occluder, which is less common in previous datasets.
Additionally, the digital scenes (with all used camera setups) and software tools we developed are made public.
More specifically, three scenes (Blender's ``\textit{Agent 327 Barbershop}'', ``\textit{Lone Monk}'', and our own novel ``\textit{\gardenname{}}'') were rendered using a novel Blender plugin, using both a cuboid- and sphere-shaped camera setup.
This resulted in the presented dataset of multi-view images and associated depth maps.
Lastly, we presented software that converts images from fish-eye projection to rectilinear projection and vice versa, to broaden applicability of the dataset.

Although the camera setups used in this paper are large compared to the related work, the Barbershop, Lone Monk and \gardenname{} scenes allow for even larger and unconventionally shaped setups.
\gardenname{} contains dynamic elements, such as the water and fire, so capturing videos instead of static images is possible.
The dataset stems from virtual environments, which allows researchers to test their approaches with ground truth depth information available (when not ambiguous due to translucency).
Future work includes extending the presented tools and software pipeline towards light field videos, which entails handling even more data and finding or creating interesting animated scenes.

\begin{acks}
This work was funded in part by the Research Foundation -- Flanders (FWO) under Grant 1SA7919N, in part by IDLab (Ghent University -- imec), in part by Flanders Innovation \& Entrepreneurship (VLAIO), and in part by the European Union.
\end{acks}

\bibliographystyle{ACM-Reference-Format}
\bibliography{main}


\end{document}